# Co-Utility: Self-Enforcing Protocols without Coordination Mechanisms

Josep Domingo-Ferrer, *Fellow*, IEEE, Jordi Soria-Comas and Oana Ciobotaru
UNESCO Chair in Data Privacy, Dept. of Comp. Eng. and Maths
Universitat Rovira i Virgili
Tarragona, Catalonia
Email: {josep.domingo,jordi.soria,oana.ciobotaru}@urv.cat

*Abstract*—Performing some task among a set of agents requires the use of some protocol that regulates the interactions between them. If those agents are rational, they may try to subvert the protocol for their own benefit, in an attempt to reach an outcome that provides greater utility. We revisit the traditional notion of self-enforcing protocols implemented using existing game-theoretic solution concepts, we describe its shortcomings in real-world applications, and we propose a new notion of self-enforcing protocols, namely co-utile protocols. The latter represent a solution concept that can be implemented without a coordination mechanism in situations when traditional self-enforcing protocols need a coordination mechanism. Co-utile protocols are preferable in decentralized systems of rational agents because of their efficiency and fairness. We illustrate the application of co-utile protocols to information technology, specifically to preserving the privacy of query profiles of database/search engine users.

## I. INTRODUCTION

Consider a set of agents who want to perform a task that requires the collaboration of all of them (like performing a computation in which each agent gives an input and expects some output, making a collective decision based on the preferences of the agents, etc.). For the task to be successfully completed, a clear set of rules that govern the expected behavior of each agent, as well as the communication between agents must be set up. We refer to such a set of rules as a protocol.

We are interested in designing protocols with respect to a given security model such that the agents taking part in them cannot subvert them. The intention of this paper is to deal with the security issues that may arise in a protocol with respect to a rational model. More precisely, we consider systems where the agents are rational and, thus, they may try to subvert the protocol, if by doing so they can increase their benefit. Our goal is to design self-enforcing protocols, or, in other words, protocols that agents are rationally interested in following.

The usual way to approach self-enforcing protocols is to use game theory, which is a mathematical tool that models the interaction between self-interested agents that act strategically. We say that an agent is self-interested when she defines a partial order over the possible outcomes of the interaction. Intuitively, one can think of this partial order as a ranking of the outcomes of the interaction. For each agent, we call this partial order the preference of the agent.

We say that an agent acts strategically when she takes into account her knowledge and her expectations about the system and about other agents before she decides on her strategy. Game theory identifies subsets of outcomes (a.k.a. solution concepts) that the agents would be most interested in achieving. In our proposal we focus on equilibrium solutions, i.e., sets of outcomes that rational agents have no intention to deviate from. Moreover, we focus our attention on the practical approaches one should take in order to implement such solution concepts.

Self-enforcing protocols are interesting because when rational agents are present, they follow them without the protocol designer having to resort to an enforcing mechanism (such as personal commitment of the agents or law enforcement systems).

In order to design self-enforcing protocols, one should start by taking into account all preferences of all agents. However, the preferences of the agents can be either public or private. Intuitively, agreeing on an equilibrium is easy if the preferences are public: each agent can compute the set of equilibria on her own (the agents may be computationally unlimited), and the protocol is then used to agree on a suitable equilibrium. Formally, this means that the outputs of the respective protocol in every run represent an equilibrium of the game. In this paper, we focus mainly on agents with private preferences. This scenario is similar to that of mechanism design, but as detailed below, we take a different approach to address it.

We consider two different types of self-enforcing protocols: those in which each agent is required to report (truthfully or not) her utility, and those that do not require agents to report her utility. We refer to the former as *coordination protocols* and to the latter as *co-utile protocols*. Both cases are interesting: coordination protocols may lead to better outcomes for all of the participants but have the difficulty of having to deal with the truthfulness of the information about preferences provided by the agents, while co-utile protocols are especially appealing because they assume only a distributed system and they achieve a good level of efficiency and fairness without agents having to report their preferences.

*Contribution and plan of this paper*. In this paper, we review self-enforcing protocols (Section II) and distinguish between those whose outcomes depend on the preferences of the agents given as inputs (Section III) and those that do not (Section IV). In connection with the latter, we introduce the notions of strict co-utility and relaxed co-utility (Section IV). In Section V, we illustrate the power of co-utile protocols for a specific application: preserving the privacy of the query profiles of database users. In Section VI, we highlight the differences between self-enforcing protocols obtained via co-utility and via mechanism design. Finally, Section VII lists some conclusions.

## II. Self-Enforcing Protocols

When performing a distributed computation in a P2P system of rational agents, each of the agents may embrace the protocol or deviate from it. Without an enforcing mechanism, a rational agent adopts the strategy that maximizes her utility. In this section we review the notion of self-enforcing protocol.

We start with an example of a protocol that is not self-enforcing: the congestion avoidance mechanism of the TCP protocol, used in Internet to avoid losing data segments in case of congestion. In plain terms, the transfer rate of TCP segments is slowly increased until a congestion is detected, which triggers a steep reduction in the transfer rate.

Consider the case of two agents Alice and Bob that are transferring data through the same router. We assume that for both players only the following two options are available: either use an implementation of the TCP protocol that properly includes the congestion avoidance mechanism or use a dishonest implementation that always transfers data at maximum rate. If both Alice and Bob use congestion avoidance, the transfer rate of both will adapt to the congestion in the network (and will be fair). If only Alice uses the implementation that includes the congestion avoidance mechanism, congestions will happen more frequently (because Bob always transfers at maximum rate) and Alice will have less bandwidth available to her. The reverse would happen if only Bob uses an implementation with the congestion avoidance mechanism. If none of them uses the congestion avoidance mechanism, the network will suffer from continuous congestion and both will get a poor service. Let us assume that the bandwidths available are those in Table I. According to the table, independently of what the other does, it is always better for oneself to use the dishonest implementation (in game theoretic terms, to be dishonest is a dominant strategy). If Alice and Bob act rationally, both will select the dishonest implementation, thereby breaking what the TCP specification says. Thus, TCP congestion avoidance is not self-enforcing. In other words, using the honest implementation of the TCP protocol is not the rational approach for agents pursuing bandwidth maximization.

TABLE I. Bandwidth available to Alice and Bob according to the selected implementation of the TCP protocol: honest (with congestion avoidance) and dishonest (without congestion avoidance)

|  |  | **Bob** | |
|---|---|---|---|
|  |  | *Honest* | *Dishonest* |
| **Alice** | *Honest* | 2,2 | 1,3 |
|  | *Dishonest* | 3,1 | 1,1 |

Self-enforcing protocols rely heavily on game theory, especially on the notion of equilibrium. An outcome is an equilibrium if no agent has incentives to change her strategy in that outcome; in other words, provided that all other agents keep their strategies unchanged, no agent can modify her strategy to increase her utility. Several notions of equilibrium have been proposed in game theory. Some of the equilibrium notions are related, being either refinements (specializations of other equilibria notions that reduce the number of solutions) or generalizations. Fig. 1 depicts the relation among several types of equilibria; arrows in the figure go from a refined to a more general equilibrium. For each notion of equilibrium there may be multiple solutions (outcomes that fulfill the requirements of the equilibrium). Which of these solutions will emerge is not determined by the solution concept and this leads to questioning if any of them will eventually emerge. By refining a notion of equilibrium, the number of solutions is reduced; in this sense, the remaining solutions become stronger. However, by using a refined equilibrium notion we may be excluding solutions that are, indeed, preferred to those that remain. When defining self-enforcing protocols it is, therefore, preferable to focus on a general notion of equilibrium (rather than on a more refined one) and rely on the protocol to enforce a specific solution.

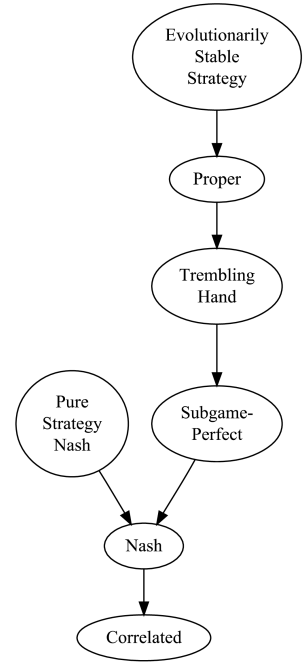

Fig 1. Graph of refinements of equilibrium concepts in game theory. The arrow goes from a refined to a more general equilibrium.

Next, we formally define self-enforcing protocols.

**Definition 1.** (Self-enforcing protocol). Let $G$ be a Bayesian game, i.e., $G = (N,A,S,T,p,U)$, where $N$ is a set of $n$

agents, $A=A_1 \times \ldots \times A_n$, $S=S_1 \times \ldots \times S_n$, $T= T_1 \times \ldots \times T_n$, $U=(u_1, \ldots, u_n)$, $A_i$ is the set of possible actions for agent $i$, $S_i$ is the set of all strategies of agent $i$[1], $T_i$ is the set of private types of agent $i$, $p:T \to [0, 1]$ is a commonly known distribution over types and $u_i:S \times T \to \mathbb{R}$ is the utility for agent $i$. Let $P:I_1 \times \ldots \times I_n \to S_1 \times \ldots \times S_n$ be a protocol taking inputs $(c_1, \ldots, c_n)$, with $c_i \in I_i$, from agents $(1,\ldots, n)$ and implementing some outcome of the game, i.e., outputting in each run a profile strategy of the game. We say that $P$ is a self-enforcing protocol for $G$ if for every input profile $(c_1, \ldots, c_n)$ and for any tuple of types $(t_1, \ldots, t_n)$, with $t_i \in T_i$, the output $P(c_1, \ldots, c_n) = (s_1, \ldots, s_n)$ is an equilibrium of the game $G$ and $u_i(s_1, \ldots, s_n, t_1, \ldots, t_n) > 0$ for all $i \in \{1, \ldots, n\}$.

Lower-bounding positive utilities by 0 in the above definition entails no loss of generality. Indeed, given a different, strictly positive lower bound that one may wish to impose on one or more utilities, one can obtain, through an affine transformation, an equivalent game $G'$ (i.e., such that there is a bijection between the equilibria in $G$ and $G'$) such that the lower bound on all utilities for $G'$ is 0. This is so because, under standard utility theory, games are insensitive (i.e., with respect to equilibria preservation) to any positive affine transformation of the payoffs.

If in Definition 1 we require that for every agent $i$ the set of private types $T_i$ have only one element, then we have the special case of a game with publicly known utilities. In this case, designing a self-enforcing protocol is just a matter of agreeing on a suitable equilibrium, e.g., by having the protocol implement only that equilibrium, which due to utilities being publicly known, anyone can compute.

For the general case, when agents do have private utilities, there are two possible approaches. In the first approach, the agents need to report some information about their utilities as input to a centralized or distributed computation that will output a strategy profile for the set of agents[2]. In the second approach, one could design a distributed protocol whose output does not depend on the agents reporting their preferences. In the following sections, we evaluate these two approaches for the design of self-enforcing protocols.

III. COORDINATION PROTOCOL

A coordination protocol is a self-enforcing protocol that takes information about the preferences of the agents as inputs. By gathering the preferences of the agents, the coordination protocol may compute the set of equilibria and then select a suitable one (thus making the protocol self-enforcing). We must keep in mind that an agent may report untruthful information about her preferences to the coordination protocol in an attempt to have it produce an outcome that is better for her.

Note that the fact that the preferences of the agents are put together to determine a suitable equilibrium does not imply the existence of a central trusted party; the computation can be distributed among the agents.

**Definition 2.** (Coordination protocol). Let $G$ be a Bayesian game and let $P: T_1 \times \ldots \times T_n \times X \to S$ be a protocol that takes as inputs the types that each agent reports together with possibly some additional information and implements some outcome of $G$. We say that $P$ is a *coordination protocol* if $P$ is a self-enforcing protocol such that $\exists \ t_1 \in T_1, \ldots, t_n \in T_n$, $\exists \ i_1, \ldots, i_k, j \in N$, $i_1 \neq j, \ldots, i_k \neq j$ and $\exists t'_{i_1} \in T_{i_1}, \ldots, t'_{i_k} \in T_{i_k}$, with $t'_{i_1} \neq t_{i_1}, \ldots, t'_{i_k} \neq t_{i_k}$, such that $u_j(P(t_1, \ldots, t'_{i_1}, \ldots, t'_{i_k}, \ldots, t_n, x), t_1, \ldots, t_n) \neq u_j(P(t_1, \ldots, t_{i_1}, \ldots, t_{i_k}, \ldots, t_n, x), t_1, \ldots, t_n)$, i.e., there exists a tuple of types and at least one agent whose utility depends on the type that one or more agents provide to the protocol.

Because the output of the coordination protocol is an equilibrium of the game, no agent has incentives to play a strategy different from the one suggested by the coordination protocol. The type of equilibrium notion is not determined by the definition. It was previously noted that it is better to rely on a general notion of equilibrium (to avoid discarding solutions that provide greater utility) and use the coordination protocol to select a specific solution. Note also that being an equilibrium is not enough. We also require the utility of each player to be greater than zero. With this condition we model the fact that users are entitled to decide between participating in the protocol or not. Assuming that a user that refrains from participating in the protocol gets zero utility, to get anyone interested in participating her expected utility must be greater than zero.

To illustrate the design of coordination protocols, we will use the following simple game. A wife and a husband are trying to decide what to do in the evening (this is the computation under consideration and the goal is to provide self-enforcing protocols for it). Each of them has to decide between going to the opera or to a football match. The wife prefers the opera, while the husband prefers the football match. This the well-known Battle of the Sexes (BoS) game [3]. Table II gives an example of actual utilities (payoffs) for husband and wife: we make the assumptions that $a_W > b_W > c$ and $a_H > b_H > c$. BoS is mainly a coordination game (both husband and wife prefer to go out together rather than on their own) but it also has elements of competitiveness (their preferences are different). In line with our previous discussion, we consider that the utilities are private: neither the wife nor the husband know the preferences of the other. The wife's view of the game is shown in Table III, where the values of $a_H, b_H, c_H$ and $d_H$ are unknown to her. The husband's view of the game is the reciprocal of Table III.

TABLE II. BATTLE OF THE SEXES (BOS) PAYOFFS.

|  |  | Husband | |
|---|---|---|---|
|  |  | Opera | Football |
| Wife | Opera | $a_W, b_H$ | $c, c$ |
|  | Football | $c, c$ | $b_W, a_H$ |

---

[1] Depending on the actual context, the set of all strategies of agent $i$ may be either her pure strategies, or her mixed strategies or another type of well-defined game theoretical strategies.

[2] Keep in mind that agents may report untruthfully, if this increases their utility.

TABLE III. WIFE'S VIEW OF THE UNDERLYING GAME.

|  |  | Husband | |
|---|---|---|---|
|  |  | Opera | Football |
| Wife | Opera | $a_W, b_H$ | $c, c_H$ |
|  | Football | $c, d_H$ | $b_W, a_H$ |

In the remainder of the section we evaluate, through examples, how the coordination protocol can be designed.

**Example 1.** In this example, we design a coordination protocol that outputs a pure strategy Nash equilibrium for a game in which the utilities of every agent do not depend on the secret types of the other agents.

Intuitively, a Nash equilibrium is a strategy profile in which each strategy is a best response to the strategies of the other agents. Thus, to determine a pure strategy Nash equilibrium it is enough for each agent to report her/his best responses to each of the possible actions of the other agent. For instance, if the wife reported truthfully, she would report "opera" as her best response to "opera".

The proposed protocol is as follows: (i) each agent reports her/his best response to each of the actions of the other player, (ii) the wife, additionally, ranks the strategies that correspond to her best responses, (iii) the set of equilibria associated to the preferences reported in (i) is computed, and (iv) the protocol selects among the computed equilibria the one with highest rank for the wife.

If each agent reported her preferences truthfully, the proposed protocol computes the set of equilibria of the underlying game. However, because the reported preferences affect the output of the protocol, agents may have incentives to be untruthful, which may lead to an outcome of the protocol that is not an equilibrium of the underlying game.

It can be shown that for the BoS game the proposed protocol makes truthful reporting a dominant strategy for the wife (because doing so yields the best equilibrium for her). A rational husband, on the contrary, may be willing to lie about his best response to "opera" if he believes that both *(opera, opera)* and *(football, football)* are equilibria of the underlying game (lying is meant to leave *(football, football)* as the only equilibrium). Therefore, the set of equilibria computed by the protocol in step (iii) is *{(opera, opera), (football, football)}* if both agents report truthfully, and *{(football, football)}* if the husband lies. The strategy profile selected in step (iv) is *(opera, opera)* and *(football, football)*, respectively. In any case, the outcome of the protocol is an equilibrium of the BoS game that gives positive utility to all the agents.

The protocol proposed in Example 1 returns a pure Nash equilibrium when BoS is the underlying game. For other underlying games this may not be the case. In general, to come up with coordination protocols we need to be able to adjust the utilities of the underlying game. This adjustment can be done via rewards and punishments, or due to the fact that the utilities used are transferable. A typical example of such a protocol is the Vickrey auction (a.k.a second-price auction) in which the price paid by the winner can be seen as a punishment. In contrast to the game described in Example 1, a Vickrey auction is a game in which the utilities of the agents may depend on the secret types of the other agents.

**Example 2.** A Vickrey auction [4] is a sealed-bid auction (each agent submits her bid without the others knowing the value). The good is assigned to the highest bidder and the price to pay is the second highest bid. If we assume that each agent has a valuation for the good, the dominant strategy for each player is to bid her true value.

The coordination protocol for the Vickrey auction is the following: gather the bids from each of the agents; the proposed strategy is the profile of collected bids. Because bidding her own valuation is a dominant strategy for each player, the proposed strategy is an equilibrium. Moreover, none of the agents gets a negative utility: except for the winner, all agents get zero utility, and the utility of the winner equals her valuation minus the second highest valuation (which results in a non-negative utility for the winner). Getting zero utility does not seem to conform to the definition of self-enforcing protocol (Definition 2), where all participants should get positive utility; however, note that participants in an auction have positive *expected utility* (because they may win).

## IV. SUPPRESSING COORDINATION MECHANISMS VIA CO-UTILITY

We have seen in the previous section that designing a coordination protocol to reach consensus on an equilibrium solution is a complex task. The main issue is that an agent may report untruthful preferences in an attempt to have the protocol produce an outcome that is preferred by her. One way to avoid this issue is to simply avoid taking agent preferences into account. The goal of this section is to design self-enforcing protocols that do not require the agents to report their preferences. We refer to such protocols as co-utile protocols.

We think of a co-utile protocol as a protocol in which a solution that is beneficial to the participating agents is achieved in a natural manner. For that, helping other agents must be the best strategy. Of course, that is only possible in a controlled environment or, in game-theoretic terms, for a specific class of agent utilities. Moreover, we require that a co-utile protocol does not need any information about the types of the participating agents, i.e., a co-utile protocol is independent of such types. In other words, we explicitly require that a co-utile protocol not be a coordination protocol.[3]

The setting we assume for co-utile protocols is suitable for sequential games. In other words, agents may take turns selecting the strategies to play. We start by delimiting the set of games that are feasible for co-utile protocols:

**Definition 3.** (Co-utility-amenable game). Let $G$ be a sequential Bayesian game for $n$ agents. We say that $G$ is a *co-utility-amenable game* if the utility of any agent is independent of the types of the other agents, i.e., $\forall i, j$, with $i \neq j$ and $\forall t_i, t_i' \in$ $T_i$ we have that $u_j(s_1, ..., s_n, t_1, ..., t_i, ..., t_n) = u_j(s_1, ..., s_n, t_1, ..., t'_i, ..., t_n)$.

---

[3] In light of Example 1, it is clear that this is not a superfluous condition: there exit protocols for co-utility amenable games that are coordination protocols.

In the rest of this paper we focus only on games with two agents. Intuitively, co-utile protocols are only feasible in a setting in which for every agent $P_i$, with $i \in \{1, 2\}$, helping the other agent is actually a rational choice. We consider the following two refinements.

We start with the definition of strictly co-utile protocols, where intuitively both agents maximize their utilities.

**Definition 4.** (Strict co-utility). Let $G$ be a co-utility-amenable game for two agents. Let $P$ be a self-enforcing protocol for $G$, such that P is not a coordination protocol. We say $P$ is a *strictly co-utile* protocol if $\forall i \in \{1, 2\}$, $\forall s'_i \in S_i$ and $\forall s'_j \in S_j$ and $\forall t_i \in T_i$ and $\forall t_j \in T_j$, with $i \neq j$, we have that $u_i(s_i, s_j, t_i, t_j) \geq u_i(s'_i, s'_j, t_i, t_j)$, where the outcome of $P$ on input $(t_i, t_j)$ is $(s_i, s_j)$.

Finally, we give the definition of relaxed co-utile protocols, where at least one agent maximizes her utility.

**Definition 5.** (Relaxed co-utility). Let $G$ be co-utility-amenable game for two agents. Let $P$ be a self-enforcing protocol for $G$, such that P is not a coordination protocol. We say $P$ is a *relaxedly co-utile* protocol if $\exists i \in \{1, 2\}$, $\forall s'_i \in S_i$ and $\forall s'_j \in S_j$ and $\forall t_i \in T_i$ and $\forall t_j \in T_j$, with $i \neq j$, we have that $u_i(s_i, s_j, t_i, t_j) \geq u_i(s'_i, s'_j, t_i, t_j)$, where the outcome of $P$ on input $(t_i, t_j)$ is $(s_i, s_j)$.

V. APPLICATION OF CO-UTILITY TO P2P ANONYMOUS QUERIES

We now give one application of co-utility to illustrate the power of this concept. The application is related to information and communication technology. In this context, if players' interests include privacy or anonymity, a protocol is co-utile if privacy preservation or anonymity preservation of an individual becomes a goal that rationally interests other individuals. We start with pairwise interactions between peer agents and then generalize to interactions between any number of agents.

*A. Pairwise interaction of agents*

Consider a system with $n$ peers $P_1, ..., P_n$ who wish to query a database or a search engine DB while keeping their interests (i.e. their query profile) private; blurring one's query profile is, intuitively, the specific notion of privacy in this example [1]. The interactions among peers take place in pairs of agents. More precisely, given two agents $P_i$ and $P_j$ that decide to interact with each other, the underlying game of their interaction is as follows[4]: the actions available to agent $P_i$, who plays the role of the initiator in this game, are either to submit by herself to the database a query $q$ she is interested in or to forward the query to $P_j$. Formally, the set of actions available to $P_i$ as initiator is $A_i = \{submit, forward\}$.

The actions available to player $P_j$ playing the role of the responder in this game are to submit the query of $P_i$ to the database or to decline the submission. Formally, the set of actions available to $P_j$ as responder is $A_j = \{submit, decline\}$.

Moreover, the utilities of any agent $P_i$ are defined as follows:

$u_i(t_i(q), f_i(q), Y_i) = \alpha_i \cdot t_i(q) \cdot f_i(q) + H(Y_i)$.

The notation[5] has the following semantics: $q$ represents a query; $t_i(q)$ represents the time left until $P_i$ would like $q$ to be answered; $f_i(q)=1$ if $P_i$ originates $q$ and is interested in getting the answer to $q$ from another agent $P_j$, and $f_i(q)=0$ otherwise ($q$ not originated by $P_i$ or $P_i$ gets the answer to $q$ directly from DB). By $H(Y_i)$ we measure the total privacy for the set of queries $Y_i$ that have been submitted to the database by $P_i$, where $H$ is the Shannon entropy. Intuitively, the higher this entropy, the flatter the histogram of frequencies of $Y_i(t)$, and the more ignorant DB stays about the query interests of $P_i$.

The parameter $\alpha_i$ describes how much $P_i$ values a quick response to her query $q$ compared to possibly reducing her amount of privacy by adding the query $q$ to her vector $Y_i$ of already submitted queries. One could, of course, compute the value of the utility $u_i$ for any query $q$, including the queries that agent $P_i$ is not interested in. However, in that case $u_i(t_i(q), f_i(q), Y_i) = H(Y_i)$, so $u_i$ is just a measure of her privacy. The interesting case arises when $q$ is a query $P_i$ is interested in but did not receive an answer for $q$ yet.

Additionally, we make the following assumptions: if an agent $P_i$ submits a query to DB, then $P_i$ will get an answer immediately. If an agent $P_i$ submits a query to another agent $P_j$, then $P_j$ is expected to accept or reject the query within a *timeout* time span (measured in seconds). In order to further simplify the presentation, we assume that, by default, $P_j$ is not interested in the answer to the query $q$, i.e., $f_j(q) = 0$.

After each action, the agents that have been influenced by that action will update their values for the functions $t, f$ and for their respective sets $Y$. For example, if $P_i$ submits $q$ directly to the database, $t_i(q)$ remains unchanged, $f_i(q)$ becomes 0 and $Y_i := Y_i \cup \{q\}$. Obviously, in this case, the utility of the partner agent $P_j$ with respect to query $q$ (i.e., $u_j(t_j(q), f_j(q), Y_j)$) remains unchanged.

In a similar way, if $P_i$ submits $q$ to $P_j$ and $P_j$ rejects the query by replying back to $P_i$ with an appropriate message, then $P_i$ updates her internal states in the following way: $t_i(q) := t_i(q) - timeout$, $f_i(q)$ remains 1 and $Y_i$ remains unchanged. Thus, the utility of $P_i$ in this case is $(t_i - timeout) \cdot \alpha_i + H(Y_i)$. For $P_j$ no internal state changes: $t_j(q) = 0, f_j(q) = 1$ and $Y_j$ remains unchanged.

Finally, if $P_i$ submits $q$ to $P_j$ and $P_j$ accepts the query by replying back to $P_i$ with the appropriate answer obtained from the database, then $P_i$ updates her internal states in the following way: $t_i(q) := t_i(q) - timeout$, $f_i(q) := 0$ and $Y_i$ stays unchanged. This means that $P_i$ has the following utility: $u_i(q) = H(Y_i)$. Similarly, $P_j$ updates her internal states in the following way: $t_j(q) = 0$ (this is unchanged), $f_j(q)$ switches to 0 and $Y_j := Y_j \cup \{q\}$, where by $Y_j \cup \{q\}$ we denote the fact that the query $q$ is appropriately added to the vector $Y_j$ of queries already submitted. Thus, $P_j$ has utility $u_j(q) = H(Y_j \cup \{q\})$.

---

4 The following game does not impose any constraints on how the decision of choosing the partner agent is made, thus giving us maximum freedom for modeling various possibilities. Both agents could decide by using a secure two-party computation that it is in their own best interest to collaborate. Or the agent who holds a query that needs to be answered may decide to randomly pick her partner and start the query game with that agent. These are just two possible settings when the interaction may arise. One could, of course, envision other possible settings.

5 In this section, we define the utility functions to depend only on the current states of the agents: In our case, the current states of an agent $P_i$ are $t_i(q), f_i(q), Y_i$ and $\alpha_i$. The current states are, in turn, computed from the initial states of $P_i$ and the partial strategies that any of the players chose so far in the game. This is just an alternative equivalent definition that we take here to simplify the presentation.

Since $P_i$ does not know in advance whether $P_j$ will answer her query or not, we make the assumption that $P_i$ believes with probability 0.5 her query will be answered. It is easy to see that the game described above is a co-utile amenable game.

For the game formally described above, we design the following protocol which we argue is an equilibrium.

**Protocol 1.** (Two-party anonymous query submission) Given a query $q$ that $P_i$ is interested in and has not yet obtained an answer for, a vector $Y_i$ of queries that $P_i$ has submitted so far to DB and a time $t_i$ that $P_i$ is willing to wait for obtaining the answer, if $u_i(t_i(q), 0, Y_i \cup \{q\}) > 1/2 \cdot (u_i(t_i(q) - timeout, 1, Y_i) + u_i(t_i(q) - timeout, 0, Y_i))$ then $P_i$ submits $q$ to DB, else $P_i$ submits $q$ to $P_j$. In turn, if $P_j$ receives a query $q$ from $P_i$, then she rejects the query if $H(Y_j) \geq H(Y_j \cup \{q\})$; otherwise, $P_j$ accepts it and answers it correctly after inquiring the database.

**Lemma.** The protocol described above is at least a relaxedly co-utile protocol for the two-player anonymous submission query game.

*Proof:* By the definition of the utilities it is easy to see that the above protocol is an equilibrium. Moreover, the players do not reveal any information about their secret types, so the above protocol is independent of them. Finally, if $H(Y_j) > 0$ and $H(Y_j \cup \{q\}) > 0$ and $H(Y_i \cup \{q\}) > 0$ and $1/2 \cdot (u_i(t_i(q) - timeout, 0, Y_i) + u_i(t_i(q) - timeout, 1, Y_i)) > 0$, then the protocol is co-utile.

If $P_i$ is strictly interested in maintaining her level of privacy by not submitting $q$ to the database and if $P_j$ is strictly interested in increasing her level of privacy by submitting $q$ to the database, then the protocol is strictly co-utile. Finally, in any other case, the protocol described above is at least relaxedly co-utile since player $P_i$'s mixed strategy ensures he obtains the highest possible utility. □

Formally, the game described above is a Bayesian game where each agent $P_i$ has an unknown type represented by $\alpha_i$ and the interest (or lack of it) for the answer to query $q$. These two values are unknown to the other agents. However, even if her type were known to the other agents, this would not influence the choices she makes. This holds true for both cases when $P_i$ plays the role of the initiator or the role of the responder.

### B. Interaction of any number of agents

Finally, let us have a look at the general case of $n$ agents participating in the system. In each of these interactions, a pair of agents aim at playing the anonymous query game described in the beginning of the section. We assume that each pair of players uses anonymous channels and the initiator agent randomly chooses her partner, i.e., the responder agent. Due to the anonymous channels assumption and also due to the utilities of the agents, any two played protocols are independent of each other: even if the agents would keep track of which other agents have helped them or not so far when asked to submit queries on their behalf (i.e., even if agents may try to hold grudge on the other agents), it is clear that the best choice for each initiator or responder is to make a choice about her next move given only her current value of the utility and the query that she wants answered or, respectively, she has just received. So, in conclusion, in our example, the interactions among the $n$ agents are, in fact, a series of independent two-party protocols.

## VI. SELF-ENFORCING PROTOCOLS AND MECHANISM DESIGN

Mechanism design is a discipline within game theory that bears some resemblance to our self-enforcing protocols proposal. The starting point in mechanism design is a set of agents that hold some private information $\theta = (\theta_1, \ldots, \theta_n) \in \Theta$, where $\Theta$ is the environment space, and a function $f$ (the social choice function) that, given a specific state of the environment $\theta$, generates the most socially desirable action. The goal is to implement the social choice function as the equilibria of a game. The diagram in Fig. 2 (extracted from [2]) illustrates this. In the diagram $M$ is the set of messages (the information that agents report; if agents report truthfully, the message is $m = \theta$, but note that agents act strategically) and $g$ is an outcome function that describes the actions taken for every $m \in M$. The pair $(M, g)$ is known as a game form, and when a state of the environment $\theta$ is realized, $(M, g, \theta)$ becomes a game. The goal is that the equilibria of the latter game be optimal according to the social choice function.

The main difference between mechanism design an self-enforcing protocols has to do with the social choice function. While the goal in mechanism design is to attain a socially desirable solution, in self-enforcing algorithms we only aim at giving each agent enough utility for her to be interested in participating. In this sense, self-enforcing algorithms have a broader field of application than mechanism design.

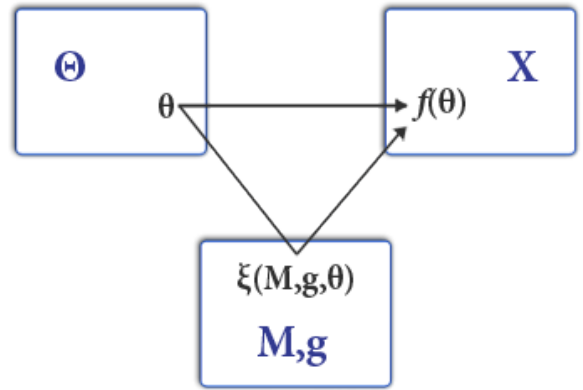

Fig 2. Given a social choice function $f$ that maps types (up-left) to a socially desired outcome (up-right), the goal of mechanism design is to come up with a game whose equilibrium implements $f$ (down).

## VII. CONCLUSIONS

We have revisited the concept of self-enforcing protocols, a class of protocols that rational agents are interested in following. We have highlighted that there are important, widely used protocols that are not self-enforcing (e.g. the TCP congestion avoidance protocol).

We have revisited one approach for the design of self-enforcing protocols which relies on a coordination mechanism that helps the agents to reach a satisfactory solution and we have defined a second approach (co-utility) that does not require the use of a coordination function. For co-utile protocols we have investigated the amount of fairness that one can achieve. This has led to two definitions of co-utile protocols: strictly co-utile protocols implement equilibria where everyone achieves her maximum utility, while in relaxedly co-utile protocols, everyone achieves at least a fair level of utility.

Beyond a theoretical discussion of self-enforcing protocols, we have also given an example of co-utile protocols that arise in information and communication technology, namely retrieving information from a database in an anonymous manner. For this example, we have defined a formal game-theoretic model of the utilities of the agents and we have provided a co-utile protocol for the anonymous query game.

Finally, we have compared self-enforcing protocols to mechanism design, a discipline in game theory that can be viewed as related. We conclude that self-enforcing protocols have a broader field of application.

ACKNOWLEDGEMENTS

The following funding sources are acknowledged: Templeton World Charity Foundation (grant TWCF0095/AB60 "Co-Utility"), Government of Catalonia (ICREA Acadèmia Prize to the second author and grant 2014 SGR 537) and Spanish Government (project TIN2011-27076-C03-01). The authors are with the UNESCO Chair in Data Privacy. The views in this paper are the authors' own and do not necessarily reflect the views of the Templeton World Charity Foundation or UNESCO.

REFERENCES

[1] J. Domingo-Ferrer and Ú González-Nicolás, "Rational behavior in peer-to-peer profile obfuscation for anonymous keyword search", *Information Sciences* 185(1):191-204, 2012.

[2] L. Hurwicz and S. Reiter, *Designing Economic Mechanisms*, Cambridge University Press, 2006.

[3] T.C. Schelling, *The Strategy of Conflict*, Harvard University Press, 1980.

[4] W. Vickrey, "Counterspeculation, auctions and competitive sealed tenders", *The Journal of Finance* 16(1):8-37, 1961.

BIOGRAPHY

**Josep Domingo-Ferrer** is a distinguished professor of computer science and an ICREA-Acadèmia researcher at Universitat Rovira i Virgili, Tarragona, Catalonia, where he holds the UNESCO Chair in Data Privacy. He received his M. Sc. and Ph. D. degrees in Computer Science from the Autonomous University of Barcelona in 1988 and 1991 (Outstanding Graduation Award). He also holds an M. Sc. in Mathematics. His research interests are in data privacy, data security, statistical disclosure control and cryptographic protocols, with a focus on the conciliation of privacy, security and functionality. He has received a Google Faculty Research Award (2014). He is a Fellow of IEEE and he has consulted worldwide on anonymization and statistical disclosure control.

**Jordi Soria-Comas** is a postdoctoral researcher at Universitat Rovira i Virgili. He has received his M. Sc. in Computer Security (2011, outstanding graduation award) and Ph. D. in Computer Science (2013) degrees from Universitat Rovira i Virgili. He also holds a M. Sc. in Finance from the Autonomous University of Barcelona (2004) and a B.Sc. in Mathematics from the University of Barcelona (2003). He also has industrial experience as financial consultant and software developer.

**Oana Ciobotaru** is a postdoctoral researcher at Universitat Rovira i Virgili. She previously served as a postdoctoral researcher at University of Trier, Germany. She has received her PhD degree in cryptography and security from Max Planck Institute for Informatics and Saarland University, Germany. She has completed a Master's degree in computer science at Max Planck Institute for Informatics and Saarland University and a Bachelor's degree in mathematics and in computer science at University of Bucharest, Romania. Her research interests include cryptography and its connections with game theory, security models with a focus on universal composability, formal methods and efficient cryptographic protocol.